\documentclass[aip,jap,amsmath,floatfix,reprint]{revtex4-1}
\usepackage{graphicx,psfrag,mathptmx}

\begin{document}

\title{Comment on ``Twin symmetry texture of energetically condensed niobium thin films on sapphire substrates (\textit{a}-plane Al$_2$O$_3$)" [J. Appl. Phys. 110, 033523 (2011)]}
\author{Paul B. Welander}\email{welander@ll.mit.edu}
\affiliation{Lincoln Laboratory, Massachusetts Institute of Technology, Lexington, MA 02420}
\date{December 14, 2011}
\begin{abstract}
In their recent publication, Zhao \textit{et al.}~[J. Appl. Phys. \textbf{110}, 033523 (2011)] claim to have found a new three-dimensional relationship for niobium-on-sapphire epitaxy.  However, two critical errors were made in the analysis of x-ray diffraction measurements.  The crystal structure of sapphire ($\alpha$-Al$_2$O$_3$) was erroneously cited as hexagonal close-packed, and crystallographic orientations of sapphire were misidentified.  Correcting these errors, one finds their claim unjustified.
\end{abstract}
\maketitle

Zhao \textit{et al.}\cite{Zhao2011}~recently reported on the growth of epitaxial niobium (110) films on A-plane sapphire ($\alpha$-Al$_2$O$_3$ (11$\bar{2}$0)) by a cathodic arc discharge technique, similar to that first used by Igarashi \textit{et al.}\cite{Igarashi1985}  This deposition method is relatively uncommon for niobium epitaxy, which is typically done by sputtering or e-beam evaporation.  Niobium epitaxy on sapphire was first achieved by Schuller over three decades ago,\cite{Schuller1980} and this metal/ceramic pair has been thoroughly investigated over the intervening years.  The extent of this body of research is so great that the topic was reviewed by Wildes \textit{et al.}~in 2001.\cite{Wildes2001}  One feature of the niobium/sapphire system that makes it unique is the three-dimensional relationship between metal and substrate, first discovered by Durbin \textit{et al.}\cite{Durbin1982} and confirmed by a number of others:\cite{Igarashi1985,Wolf1986,Nishihata1986,Claassen1987,Mayer1990a,Yoshii1995,Wagner1996}
\begin{center}
  Al$_2$O$_3$ [11$\bar{2}$0] $\parallel$ Nb [110] and Al$_2$O$_3$ [0001] $\parallel$ Nb [1$\bar{1}$1]
\end{center}
This relationship is by no means universally observed -- under certain growth and annealing conditions other orientational relationships have been found.\cite{Oya1986,Wagner1996,Wagner1998,Dietrich2003}  However, Zhao \textit{et al.}~go one step further, calling into question the validity of this relationship and instead proposing their own three-dimensional registry.  My comment begins with a summary of their arguments.

Utilizing x-ray diffraction (XRD) measurements -- pole figures and $\varphi$ scans, specifically -- Zhao \textit{et al.}~observe crystallographic twins in all but one of their niobium films (see their Fig.~2).  The one exception is a film deposited at the highest reported growth temperature (400~$^{\circ}$C) that also shows the highest residual resistance ratio.  To explain the existence of these twins in their poorer quality films, they note that the six-fold symmetry of sapphire about the (0001), or C-, axis and the three-fold symmetry of niobium about the (111) axis (see their Fig.~4) allows for \textit{two equivalent orientations} of Nb (110) on A-plane sapphire (see their Fig.~5).  Furthermore, and despite the fact that their niobium films on A-plane sapphire are all (110)-oriented, Zhao \textit{et al.}~argue that the three-dimensional registry should have Al$_2$O$_3$ [10$\bar1$0] $\parallel$ Nb [110] instead.  One feature of their XRD data that the authors admit to having no explanation for is the fact that in all of their twinned films, one twin is always dominant (the two do not appear equivalent).

\begin{figure}[b]
  \includegraphics[width=3.000in]{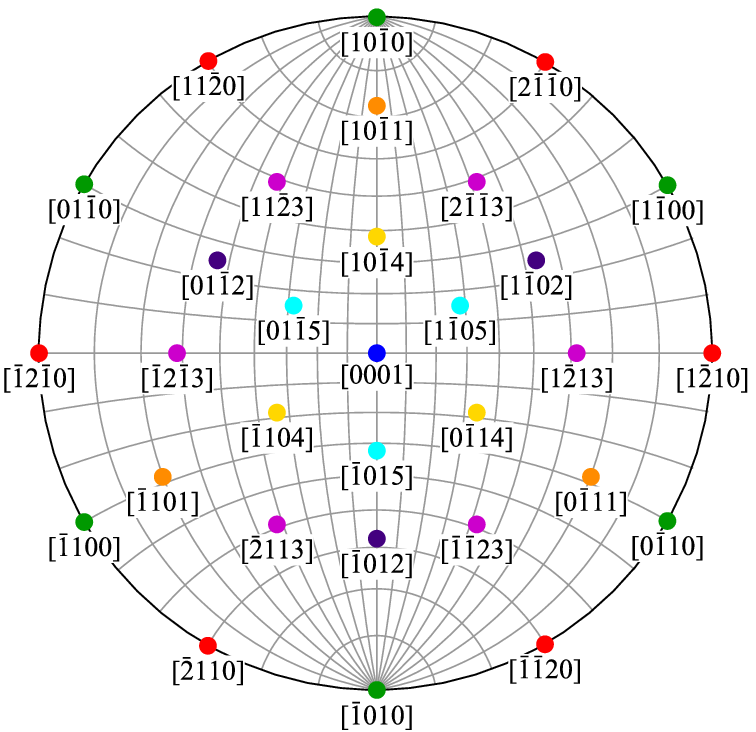}
  \caption{(Color online) A (0001) stereographic pole plot of $\alpha$-Al$_2$O$_3$  showing the three-fold symmetry of the crystal lattice.}
  \label{Figure1}
\end{figure}

\begin{figure}[t]
  \includegraphics[width=3.375in]{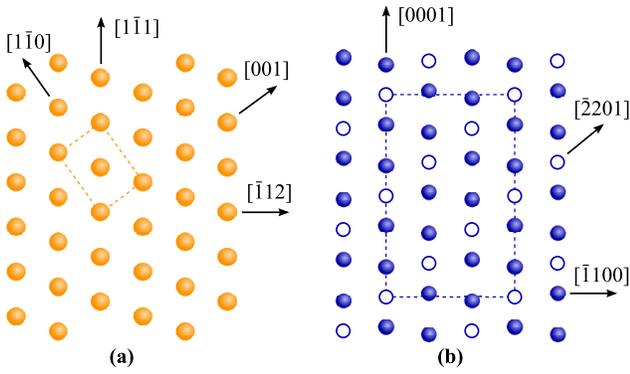}
  \caption{(Color online) Atomic arrangements of (a) the bcc Nb (110) plane and (b) the Al$^{3+}$ sublattice in $\alpha$-Al$_2$O$_3$ (11$\bar{2}$0) (with open circles denoting vacancy positions).  Dashed lines denote the unit cells.}
  \label{Figure2}
\end{figure}

However, in analyzing their niobium films the authors make two critical errors.  The first of these concerns the crystal structure of sapphire.  Zhao \textit{et al.}~erroneously state that sapphire (the corundum form of Al$_2$O$_3$, often denoted by the Greek letter $\alpha$) has a hexagonal close-packed (hcp) lattice.  In fact, $\alpha$-Al$_2$O$_3$ has a much more complicated structure that has been described in detail by Kronberg\cite{Kronberg1957} and Lee \textit{et al.}\cite{Lee1985} --  the oxygen anions occupy an hcp lattice, with aluminum cations filling two thirds of the octahedral interstices.  The aluminum vacancies give rise to distortions in both the anion and cation sublattices, and their ordering is what defines the sapphire unit cell.  The end result is a hexagonal structural unit cell with three-fold symmetry about the C-axis (see Fig.~\ref{Figure1}), not the six-fold symmetry one finds with a pure hcp crystal structure.  This reduced symmetry undermines the argument of Zhao \textit{et al.}~in favor of two equivalent orientations of (110) niobium on A-plane sapphire.  In fact, the apparent symmetry of the A-plane surface is broken by the arrangement of aluminum vacancies (see Fig.~\ref{Figure2}) -- the twins described by Zhao \textit{et al.}~are non-equivalent for this reason.

The second error concerns the misidentification of crystal axes in the sapphire basal plane.  In their Fig.~4, Zhao \textit{et al.}~show a top view of the (0001) sapphire lattice, with arrows indicating both the [01$\bar1$0] (thin, black) and [10$\bar1$0] (thick, green) directions.  These two arrows are actually drawn along two $\langle$11$\bar2$0$\rangle$ axes instead.  Consider the correctly drawn (11$\bar2$0) plane (red line) in the same figure -- the normal to that plane is the [11$\bar2$0] direction, and is equivalent to the axes indicated by the arrows.  This also means that both of the associated pole figures need to be rotated 30$^{\circ}$ for consistency.  With this correction, the relationship first established by Durbin \textit{et al.}\cite{Durbin1982} is confirmed, with Al$_2$O$_3$ [11$\bar2$0] $\parallel$ Nb [110].


This work was sponsored by the Department of the Air Force under Contract Number FA8721-05-C-0002.  Opinions, interpretations, conclusions, and recommendations are those of the authors, and not necessarily endorsed by the United States Government.


%

\end{document}